# PHOTOCURRENT ENHANCEMENT OF SPIN COATED CDS THIN FILMS BY ADDING CU

[1]P. Samarasekara, [1]B.M.M.B. Basnayaka and [2]Sunil Dehipawala

[1]Department of Physics, University of Peradeniya, Peradeniya, Sri Lanka

[2] Physics Department, Queensborough Community College of CUNY, 222-05, 56th Avenue, Bayside, NY 11364, USA

*Abstract*

*$Cu^{2+}$ added CdS films were synthesized using spin coating technique at different spin speeds for different time durations. Films were subsequently annealed at different temperatures for different time periods in air to crystallize the phase of CdS in thin film form. Films were characterized using XRD, UV- visible spectrometer and solar simulator. According to XRD patterns, addition of trace amount of Cu didn't change the structure of CdS. However, the optical band gap gradually decreases with percentage of Cu as expected. As a result, the photocurrent, fill factor and efficiency measured in $KI/I_2$ electrolyte gradually increase with the amount of Cu. Photovoltaic properties could be improved without altering the structure of CdS. Efficiency enhanced CdS films find potential applications in solar cell industry.*

*Keywords:* *CdS, Cu doping, optical band gap, photovoltaic properties, Tauc model*

## 1. Introduction:

CdS is n-type wide band gap material with band gap of 2.42eV. CdS find potential applications in window materials in solar cells, light emitting diodes, sensors, photoconductors, optical mass memories and solar selective coatings [1]. The structure of Cadmium Sulfide (CdS) is solid hexagonal or cubic crystal. CdS films have been deposited using many different techniques. Highly luminescent Cu doped CdS nanorods were synthesized using spin coating technique, and the optical properties of them have been investigated [2]. P-type Cu doped CdS films have been fabricated, and Seebeck and Hall coefficients of these films have been determined [3]. Photocurrent due to carrier combination in Cu doped CdS thin films prepared using spray pyrolysis technique has been measured [4]. X-ray photoelectron spectroscopy (XPS) analysis of p-type Cu doped CdS thin films has been performed [5]. Chemical states and depth profiles of Cu, Cd and S in p-type Cu doped CdS thin films have been investigated using Auger parameter [5]. Undoped and Cu doped CdS thin films were characterized using photothermal deflection spectroscopy (PDS) and X-ray diffraction (XRD) techniques [6]. Changes of carrier mobility, grain size and lattice strain due to the addition of Cu have been investigated [6]. Structural and photoelectrochemical properties of Cu doped CdS thin films synthesized using ultrasonic spray pyrolysis have been investigated [7]. Optical and structural properties of Cu doped CdS thin films prepared by pulsed laser deposition have been studied [8].

Pure CdS thin films have been previously fabricated using spin coating by us [9]. In addition, screen printed pure CdS films have been characterized [10]. The structural properties of pure CdS deposited by chemical bath deposition and pulsed direct current magnetron sputtering have been reported [11]. Influence of sol aging time and annealing temperature on physical properties of pure nanocrystalline CdS thin films have been investigated [12]. Pure CdS thin films have been fabricated by electrodeposition, and effect of colloidal sulfur particle stability on film composition has been investigated [13]. Properties of nanocrystalline pure CdS thin films deposited by spray pyrolysis have been studied [14].

Previously thin films of ZnO, p-$Cu_2O$/n-CuO, carbon nanotubes, lithium mixed ferrite and CuO films have been synthesized using reactive DC sputtering, rf sputtering, chemical vapor deposition (CVD) techniques by us [15, 16, 17, 18, 19]. A vacuum is incorporated in all above sputtering and CVD techniques. However, spin coating technique is considered to be a low cost film fabrication method compared to the technique required vacuum to deposit films. In addition, the band gap variation of semiconductor particles doped with salts was investigated by us [20]. CdS retains some magnetic properties [21]. Magnetic properties of thin films have been theoretically explained using Heisenberg Hamiltonian by us [22, 23, 24]. Photovolataic properties of thin films can be enhanced by coating dye layers as well [25]. However, the photovoltaic properties enhanced of CdS by adding Cu is presented in this mnauscript.

## 2. Experimental:

First, cupric acetate and Ethanol were mixed together to make a stock solution of $Cu^{+2}$ with a concentration of 0.02 moldm$^{-3}$. Then, this solution was continuously stirred for 24 hours. Separately, 3% of PEG (w/W%) was dissolved in 2.5ml of ethanol, and 0.145g of Cadmium acetate was added to the solution. In the same way, 0.095g of Thiourea was dissolved in 2.5ml of ethanol and both solutions were stirred separately for a one hour using a magnetic stirrer. Then cadmium acetate solution and Thiourea soultion were mixed together. Two drops of Nitric acid were added to the final solution as a catalyst. The $Cd^{2+}$ concentration of the resulting solution was equal to 0.5moldm$^{-3}$ and the molar ratio of Cd:S was 1:2. This solution was stirred again for another 24hours and filtered. Finally, the solution was kept for 2 days for aging at room temperature. Different calculated volumes from the $Cu^{2+}$ stock solution were added to the CdS solution to obtain Cu: Cd ratios of 0.1%, 0.2%, 0.3%, 0.4%, 0.5% and 0.6%. Then pure CdS and Cu added CdS thin films were deposited on ultrasonically cleaned glass substrates by sol-gel spin-coating technique. Solution was dropped onto the glass substrates at speeds of 1500, 2200 and 2400 rpm for 30 seconds. After deposition, the film samples were dried on hot plate at 120 $^0$C for 10 minutes and then annealed at temperatures of 200 $^0$C, 300 $^0$C, and 400 $^0$C in air for one hour.

Structural properties of film samples were determined using X ray diffraction (XRD) with Cu-$K_\alpha$ radiation of wavelength 1.54060 Å. UV-visible spectrometer was employed to investigate optical properties of samples. Photocurrent and photovoltage were measured using solar cell simulation system PET cell L01. Electrolyte used to measure photovoltaic properties was KI/$I_2$. Films grown on normal insulating glass substrates were used for XRD and optical property measurements. Films deposited on conductive glass (ITO) substrates were used for photocurrent and photovoltage measurements.

## 3. Results and Discussion:

When the $Cu^{2+}$ concentration was increased, the follow-on thin films appeared to be patchy without having a uniform distribution. Moreover, the color of the thin film became brownish with the increment of the $Cu^{2+}$ concentration. Cu can easily defuse into the CdS, since Cu has s diffusion coefficient of $10^{-12}$ cm$^2$s$^{-1}$ which is quiet a large value at room temperature. Cu can also be substituted or it could occupy a Cd vacancy to create a deep acceptor state. Figure 1 shows the XRD pattern of pure CdS thin film. The XRD pattern of Cu doped CdS is similar to the XRD pattern of pure CdS by indicating that the addition of trace amount of Cu doesn't change the structure. This can happen due to different reasons as follows. The phase of Cu in the film can be amorphous. Or if the Cu is crystallized at the grain boundaries, XRD pattern will not detect Cu. Or because the added amount of Cu is very small, the peaks of Cu can be weak compared to the peaks of CdS. Or if the Cu atoms occupy the vacant sites in the CdS lattice, then the lattice structure will remain the same.

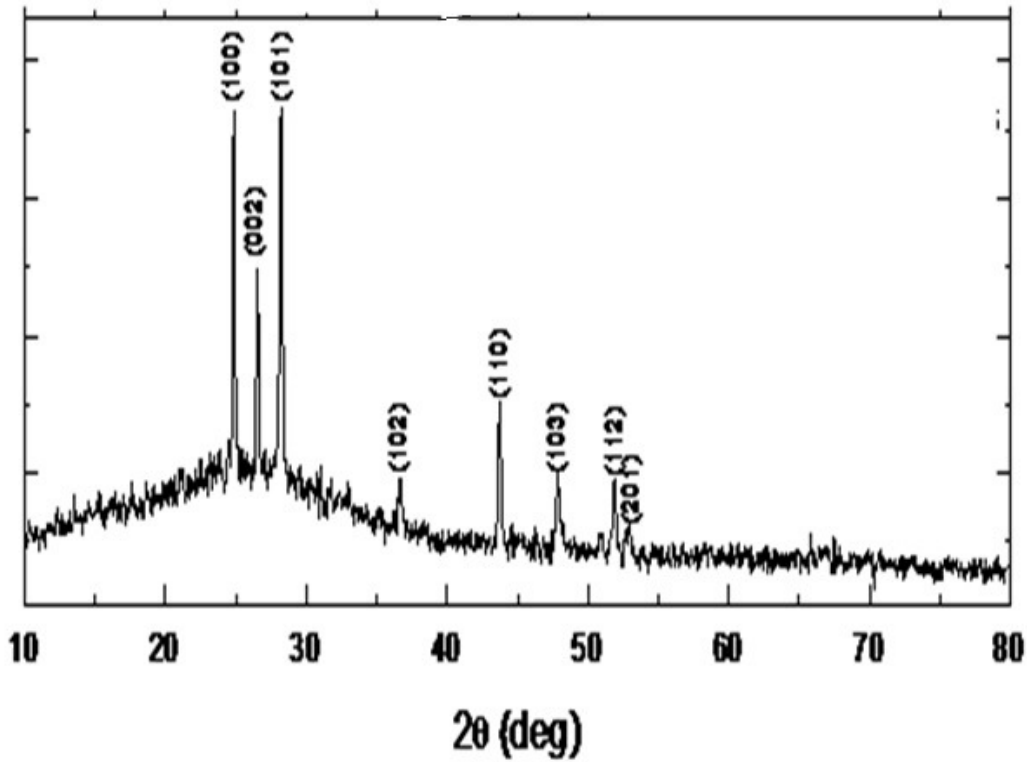

Figure 1: XRD pattern of spin coated pure CdS thin films.

First spin speed, spin time, annealing time, annealing temperature were varied to find the optimum conditions. All the samples described in this section were spin coated at spin speed of 900 rpm for 30 seconds, and subsequently annealed at 300 $^0$C for 20 minutes. The average crystalline size (D), lattice parameters (a, c), strain (ε) and dislocation density (δ) of pure CdS sample and Cu doped CdS sample were calculated using XRD patterns [9]. The Crystalline size (D) of pure CdS and Cu 0.3% doped films were calculated using the Scherrer formula. The lattice parameters reduce with addition of Cu. As a result, the average crystalline size decreases. In addition, the strain and dislocation density increase with the addition of Cu. The deformation taking place due to the addition of an impurity (Cu) is the reason for the increase of strain and dislocation density.

The optical properties of samples were measured by means of UV-visible spectroscopy. When $I$ and $I_0$ are the measured intensities of outgoing and incoming waves, respectively, the transmittance ($T$) is given by $T\% = \frac{I}{I_0} \times 100\%$. Then absorbance ($A$) was calculated using $A = -\log_{10}(T)$. Finally the absorption coefficient (α) was calculated using $\alpha = 2.303 \times \left(\frac{A}{t}\right)$ at each value of frequency (ν). Here $t$ is the thickness of the film sample. Then the graph of $(\alpha h\nu)^2$ versus hν, called Tauc plot, was plotted for each sample. Here $h$ is the Plank's constant. The relationship $\alpha h\nu = B(h\nu - E_g)^{0.5}$ was used to calculate the optical band gap. When the value of $\alpha h\nu$ is zero on the graph, the optical band gap is given by $E_g = h\nu$. Figure 2 and 3 show the Tauc plot for 0.1% Cu doped and 0.6% Cu doped CdS samples, respectively. Table 1 show the variation of optical band gap with $Cu^{2+}$ doping concentration. The optical band gap decreases with the doping concentration due to the addition of extra energy levels. The optical band gap at 0.1% Cu doping concentration is slightly higher than that of bulk CdS (2.42eV). The optical band gap at all other Cu doping concentrations is less than that of bulk CdS.

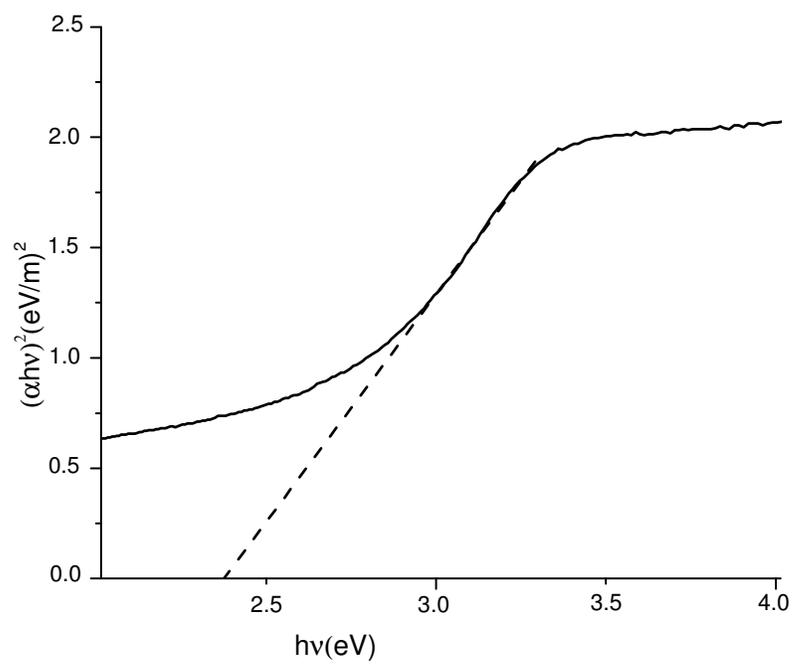

Figure 2: Graph of $(\alpha h\upsilon)^2$ versus $h\upsilon$ for 0.1% Cu doped CdS thin film.

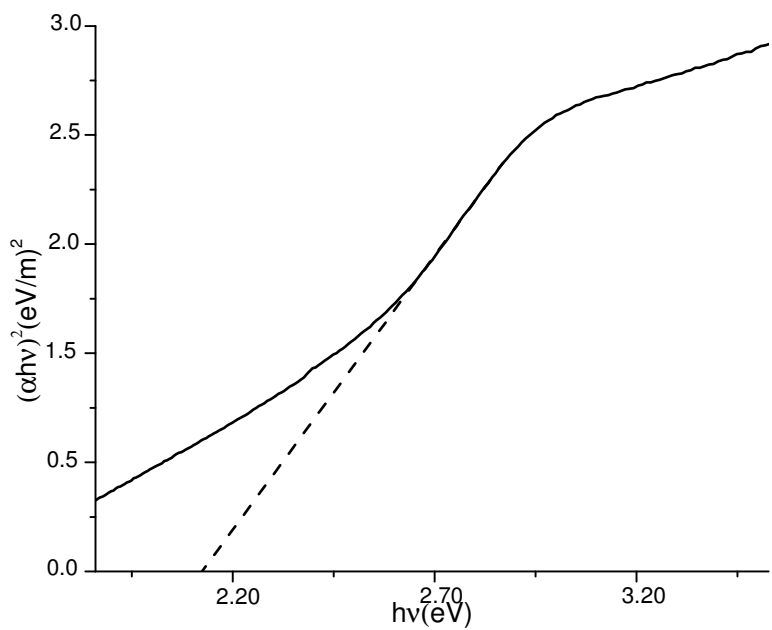

Figure 3: Graph of $(\alpha h\upsilon)^2$ versus $h\upsilon$ for 0.6% Cu doped CdS thin film.

| $Cu^{2+}$ Concentration (%) | Optical Band Gap Energy (eV) |
|---|---|
| 0.1 | 2.45 |
| 0.2 | 2.39 |
| 0.3 | 2.27 |
| 0.4 | 2.28 |
| 0.6 | 2.18 |

Table1: Optical band gap at different $Cu^{2+}$ doping concentrations.

Photo current - photo voltage experiment was performed under visible light irradiation by blocking UV irradiation using a wavelength cutoff. Solution of KI/$I_2$ was used as the electrolyte. Current –Voltage characteristics was plotted for different Cu impurity percentages in the CdS thin films as given in figure 4. Only the curves for CdS doped with 0.1% and 0.2% are given in figure 4 for the clarity. However, the I-V characteristics were plotted for other Cu concentrations as well. Open circuit voltage ($V_{OC}$), short circuit current density ($J_{SC}$), fill factor (FF) and efficiency were calculated using the data given in figure 4 as shown in table 2.

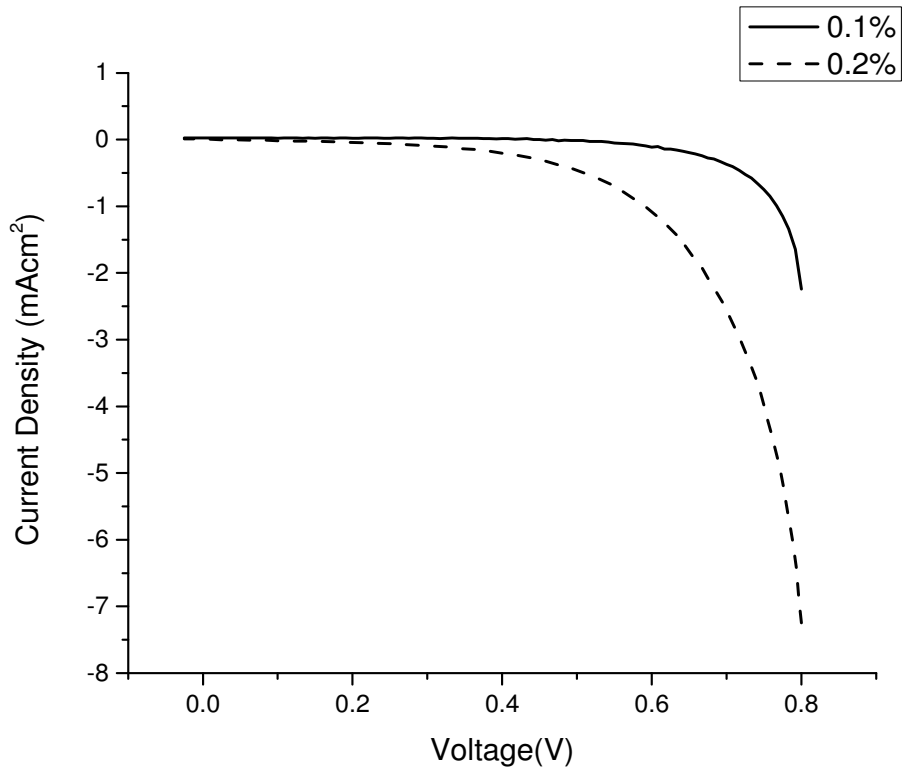

Figure 4: Current density versus voltage for two doping concentrations of $Cu^{2+}$.

| Cu Concentration | $V_{OC}$(V) | $J_{SC}$(mAcm$^{-2}$) | Fill Factor (FF%) | Efficiency (%) |
|---|---|---|---|---|
| 0.1 | 0.3259 | 0.013 | 19.69 | 0.09 |
| 0.2 | 0.2719 | 0.025 | 21.21 | 0.21 |
| 0.3 | 0.1599 | 0.03 | 27.10 | 0.44 |
| 0.4 | 0.1497 | 0.035 | 31.09 | 0.44 |
| 0.5 | 0.0781 | 0.039 | 43.87 | 0.63 |
| 0.6 | 0.0498 | 0.046 | 49.21 | 0.86 |

Table 2: Photovoltaic data for samples with different Cu $^{2+}$ concentrations.

According to table 2, the photocurrent density increases with the doping concentration due to the decrease of optical band gap. When optical band gap is smaller, more electrons are allowed to jump to higher energy levels. As a result of increase of photocurrent, both the fill factor and efficiency increase with the Cu concentration.

## 4. Conclusion:

The structure of CdS didn't change due to the doping of trace amount of $Cu^{2+}$. Reasons can be attributed to the amorphous Cu phase, crystallization of Cu at grain boundaries or weak XRD peaks of trace amount of Cu. Due to addition of Cu to the CdS film, the lattice parameters slightly reduces, particle size decreases, strain increases, and dislocation density increases. Optical band gap gradually decreases with the $Cu^{2+}$ doping concentration from 2.45 to 2.18eV due to the addition of impurity levels. However, the optical band gap of Cu doped film samples varies around the optical band gap of bulk CdS (2.42eV). As a result, photocurrent density increases from 0.013 to 0.046 mA/cm$^{-2}$, fill factor increases from 19.69 to 49.21%, and efficiency increases from 0.09 to 0.86% with doping concentration. The efficiency of a solar cell can be increased approximately by a factor of 10 by adding a trace amount of Cu. Because the photocurrent density, fill factor and efficiency increase with the addition of Cu, the Cu doped samples will make some developments in solar cell industry.

**References:**

1. Olopade M.A., Awobode A.M., Awe O.E. and Imalerio T.I., 2013. Structural and optical characterization of sol-gel spin- coated nanocrystalline CdS thin film. *International journal of research and reviews in applied sciences* 15(1), 120-124.
2. Kumar N., Purohit L.P. and Goswami Y.C., 2015. Spin coating of highly luminescent Cu doped CdS nanorods and their optical structural characterizations. *Chalcogenide letters* 12(6), 333-338.
3. Kashiwaba Yasube, Kanno Itaru and Ikeda Toshio, 1992. P-type characteristics of Cu doped CdS thin films. *Japanese Journal of Applied Physics* 31(1), 1170-1175.
4. Panda Richa, Rathore Vandana, Rathore Manoj, Shelke Vilas, Badera Nitu, Sharath Chandra L.S., Jain Deepti, Gangrade Mohan, Shripati T. and Ganesan V., 2012. Carrier recombination in Cu doped CdS thin films: Photocurrent and optical studies. *Applied surface science* 258(12), 5086-5093.
5. Abe Takashi, Kashiwaba Yasube, Baba Mamoru, Imai Jun and Sasaki Hideyuki, 2001. XPS analysis of p-type Cu doped CdS thin films. *Applied surface science* 175-176, 549-554.
6. Paulraj M., Ramkumar S., Varkey K.P., Vijayakumar K.P., Sudha Kartha C. and Nair K.G.M., 2005. Characterizations of undoped and Cu doped CdS thin films using photothermal and other techniques. *Physica Status Solidi a* 202(3), 425-434.
7. Xie Rui, Su Jinzhan, Li Mingtao and Guo Liejin, 2013. Structural and photoelectrochemical properties of Cu doped CdS thin films prepared by ultrasonic spray pyrolysis. *International journal of photoenergy* http://dx.doi.org/10.1155/2013/620134.
8. Mahdavi S.M., Irajizad A., Azarian A. and Tilaki R.M., 2008. Optical and structural properties of copper doped CdS thin films prepared by pulsed laser deposition. *Scientia Iranica* 15(3), 360-365.
9. Dissanayake D.M.C.U. and Samarasekara P., 2015. Optical and structural Properties of Spin Coated Cadmium Sulphide Thin Films. *Journal of science: University of Kelaniya* 10, 13-20.

10. Kumar V., Sharma D.K., Bansal M.K., Dwivedi D.K. and Sharma T.P., 2011. Synthesis and characterization of screen printed CdS films. *Science of sintering* 43(3), 335-341.

11. Lisco P., Kaminski P.M., Abbas A., Bass K., Bowers J.W., Claudio G., Losurdo M. and Walls J.M., 2015. The structural properties of CdS deposited by chemical bath deposition and pulsed direct current magnetron sputtering. *Thin solid films* 582, 323-327.

12. Rathinamala I., Pandiarajan J., Jeyakumaran N. and Prithivikumaran N., 2014. Synthesis and physical properties of nanocrystalline CdS thin films- Influence of sol aging time and annealing temperature. *International Journal of Thin Films science and Technology* 3 (3), 113-120.

13. Takahashi M., Hasegawa S., Watanabe M., Miyuki T., Ikeda S. and Iida K., 2002. Preparation of CdS thin films by electrodeposition: effect of colloidal sulfur particle stability on film composition. *Journal of applied electrochemistry* 32, 359-367.

14. Yadav A.A., Barote M.A. and Masumdar E.U., 2010. Studies on nanocrystalline cadmium sulphide (CdS) thin films deposited by spray pyrolysis. *Solid sate sciences* 12(7), 1173-1177.

15. Samarasekara P., Nisantha A.G.K. and Disanayake A.S., 2002. High Photo-Voltage Zinc Oxide Thin Films Deposited by DC Sputtering. *Chinese Journal of Physics* 40(2), 196-199.

16. Samarasekara P., 2010. Characterization of Low Cost p-$Cu_2$O/n-CuO Junction. *Georgian electronic scientific journals: Physics* 2(4), 3-8.

17. Samarasekara P., 2009. Hydrogen and Methane Gas Sensors Synthesis of Multi-Walled Carbon Nanotubes. *Chinese Journal of Physics* 47(3), 361-369.

18. Samarasekara P., 2002. Easy Axis Oriented Lithium Mixed Ferrite Films Deposited by the PLD Method. *Chinese Journal of Physics* 40(6), 631-636.

19. Samarasekara P. and Yapa N.U.S., 2007. Effect of sputtering conditions on the gas sensitivity of Copper Oxide thin films. *Sri Lankan Journal of Physics* 8, 21-27.

20. Tennakone K., Wickramanayake S.W.M.S., Samarasekara P. and Fernando C.A.N., 1987. Doping of Semiconductor Particles with Salts. *Physica Status Solidi (a)* 104, K57-K60.

21. Zhao X.G., Chu J.H. and Tang Z., 2015. Magnetic properties, Heisenberg exchange interaction, and curie temperature of CdS nanoclusters. *The journal of physical chemistry*

119(52), 29075-29086.
22. Samarasekara P. and Saparamadu Udara, 2013. Easy axis orientation of Barium hexa-ferrite films as explained by spin reorientation. *Georgian electronic scientific journals: Physics* 1(9), 10-15.
23. Samarasekara P. and Saparamadu Udara, 2012. Investigation of Spin Reorientation in Nickel Ferrite Films. *Georgian electronic scientific journals: Physics* 1(7), 15-20.
24. Samarasekara P. and Gunawardhane N.H.P.M., 2011. Explanation of easy axis orientation of ferromagnetic films using Heisenberg Hamiltonian. *Georgian electronic scientific journals: Physics* 2(6), 62-69.
25. Kumara G.R.A., Tiskumara J.K., Ranasinghe C.S.K., Rathnayake I.S., Wanninayake W.M.N.M.B., Jayaweera E.N., Bandara L.R.A.K. and Rajapakse R.M.G., 2013. Efficient solid state dye sensitized n-ZnO/D-358dye/p-CuI solar cell. *Electrochimica Acta* 94, 34-37.
12